\documentclass[twocolumn,preprintnumbers,amsmath,floatfix,prb,aps,superscriptaddress]{revtex4-1}

\usepackage{graphicx}
\makeatletter
\usepackage{amssymb}

\begin{document} 

\title{Chiral-mediated entanglement in an Aharonov-Bohm ring}

\author{Bruno Rizzo} \affiliation{Departamento de F\a'{i}sica,
  FCEyN and IFIBA, Universidad de Buenos Aires, Pabell\'on 1, Ciudad
  Universitaria, 1428 Buenos Aires, Argentina}

\author{Liliana Arrachea} \affiliation{Departamento de F\a'{i}sica,
  FCEyN and IFIBA, Universidad de Buenos Aires, Pabell\'on 1, Ciudad
  Universitaria, 1428 Buenos Aires, Argentina}

\author{Juan Pablo Paz} \affiliation{Departamento de F\a'{i}sica,
  FCEyN and IFIBA, Universidad de Buenos Aires, Pabell\'on 1, Ciudad
  Universitaria, 1428 Buenos Aires, Argentina}

\date{\today}

\begin{abstract}

We study the  orbital entanglement in a biased  Aharonov-Bohm ring
connected in a four-terminal setup. We find that the concurrence 
achieves a maximum when the magnetic flux $\Phi_B$ coincides with an
integer number of half a flux quantum $\Phi_0/2$. We show that this 
behavior is a consequence of the existence of degenerate states 
of the ring having opposite chirality. We also analyze the 
behavior of the noise as a function of $\Phi$ and discuss the 
reliability of this quantity as an entanglement witness.

\end{abstract}


\maketitle

\
\section{INTRODUCTION}
The increasing interest in quantum information processing is 
boosting the search for mechanisms to produce and control
entanglement in devices of different nature. \cite{qe} Photonic
devices are routinely used for preparing and detecting entangled photon
pairs.\cite{inter} The main limitations of such devices is the non-existence
of deterministic sources as well as the difficulty in controlling the 
interaction between the photons. Entanglement mechanisms have also been 
proposed in solid state devices, like quantum dots, \cite{entqdot}  
and superconductors.\cite{entsup}  The fundamental ingredient 
behind all these mechanisms is a many-body interaction. More recently, it was determined
that entanglement is also possible in systems of non-interacting 
electrons.\cite{been1, been2} In particular, it was shown that electron-hole entangled 
pairs can be produced by biasing a tunneling barrier.  

The edge states of systems in the quantum Hall regime can be employed 
in solid state devices to produce electron beams with properties similar
to those of a photon beam in optical setups. 
Exploiting this analogy, several theoretical and experimental proposals 
of electronic interferometers have been reported. \cite{elin}  
Interestingly, the electronic counterpart of the Hanbury-Brown-Twiss
device has been analyzed in a configuration of  edge Hall states that do not have 
interfering orbits. \cite{hbt,but1}  
For this reason, Aharonov-Bohm  (AB) effect takes 
place only at the level of two-particle correlation functions, 
while the single-particle AB effect is not present. 
This entanglement seems to be related with an asymmetry in the 
device which favors the internal production of particle-hole 
pairs and manifest itself in the behavior of the current-current correlation functions.   
The relevance of the current-current correlation functions as entanglement
witness has been also discussed in other fermionic systems. \cite{noise-ent}

The aim of this work is to establish the existence of orbital 
entanglement in AB systems where single-particle 
interference is present. Using a microscopic model, 
we show that when coupling the AB-ring to four leads (two on the right
and two on the left), the post select two-particle states of electrons at opposite 
leads are typically 
entangled. We name it {\em ''chiral-mediated entanglement''} (CME) because its creation  is possible by the existence of intermediate 
states in the AB-ring which are coherent superpositions
of two different chiralities for the electronic motion. Remarkably, this kind of entanglement can be also defined in 
a transport setup, which is identical to the one that was proposed 
to define the electron-hole entanglement, \cite{been1,been2,but1,but2,frust1,frust2}. 
In fact, we also show here that the noise current-current 
correlation function can be a good witness for the entanglement in our
setup. 

The paper is organized as follows. In Sec. II, we present
the model in detail and we sketch the formulation of the theory,
where we calculate the reduced density-matrix of a two-particle system. 
The results of this work are presented in Sec. III. The conclusions are presented
in Sec. IV.
\
\section{THEORETICAL TREATMENT}

\subsection{Model}

We consider the AB single-channel ring with
a magnetic flux. The ring is connected to four terminals 
\cite{been1,been2,but1,but2,frust1} as shown in Fig. \ref{fig1}. 
Two of the terminals, those labeled by $\alpha=1,2$, at the 
left side of the ring, are at a higher voltage $V$ with 
respect to the ones at the right, labeled 
by $\alpha=3,4$. All terminals are ordinary single-channel 
metallic leads where electrons can move either to the right or to 
the left. For simplicity, we consider spinless 
electrons and we describe the setup by the Hamiltonian 
\begin{equation} \label{ham}
H=\sum_{\alpha=1}^4 \left( H_{\alpha} + H_{c,\alpha} \right) + H_{ring},
\end{equation}
where 
$H_{\alpha} = \sum_{k_{\alpha}} \varepsilon_{k_{\alpha}} 
c^{\dagger}_{k_{\alpha}} c_{k_{\alpha}}$ are Hamiltonians of non-interacting electrons 
representing the
leads. For the AB-ring we use a non-interacting  model, where 
electrons move with velocity $v$ either clockwise ($+$ chirality) 
or anticlockwise ($-$ chirality). The Hamiltonian is 
\begin{equation}\label{ring}
 H_{ring} = \sum_{\lambda=\pm}  \int_0^L dx    v \lambda 
\Psi^{\dagger}_{\lambda}(x) {\cal D}_x \Psi_{\lambda}(x) ,
\end{equation}
being $\lambda$ the chirality, $ {\cal D}_x = - i \partial_x -  \phi, \;\; \phi= \Phi/(L \Phi_0)$, 
with $\Phi=2 \pi \Phi_B$, where $\Phi_B$ is the magnetic flux, $\Phi_0=h c/e$ is the flux quantum 
and $L$ is the length of the ring. 
The contacts between the leads and the ring are modeled by
tunneling terms of the form 
$H_{c \alpha} = \sum_{k_{\alpha}, \lambda=\pm } w_{k_{\alpha}} [c^{\dagger}_{k_{\alpha}} 
\Psi_{\lambda}(x_{\alpha})+ H.c.]$, 
where $x_{\alpha}$ define the positions of the ring to which the 
leads are attached. We consider the leads to be at zero temperature. 
The chemical potentials enforce a bias voltage between 
left and right leads, i.e. $\mu_1=\mu_2=\mu_L$; 
$\mu_3=\mu_4= \mu_R$ and $\mu_L-\mu_R= eV$, which we assume to be very small
$eV \sim 0$. 
\
\subsection{Reduced Density Matrix and Concurrence}

In a setup as the one in Fig. \ref{fig1}, electrons tunnel from the 
left leads to the right ones. We aim to define the effective 
density matrix describing the quantum state post-selected from the total 
two--electron state by projecting out the components where both electrons are 
either in the right or in the left leads. \cite{been2,but2,frust1} 
We introduce the operators $A^\dagger_{00} \equiv c_{k_{1}}^{\dagger}c_{k_{3}}^{\dagger},
 A^\dagger _{01} \equiv c_{k_{1}}^{\dagger}c_{k_{4}}^{\dagger},
 A^\dagger _{10} \equiv c_{k_{2}}^{\dagger}c_{k_{3}}^{\dagger},
 A^\dagger _{11} \equiv c_{k_{2}}^{\dagger}c_{k_{4}}^{\dagger} $, which 
 create one particle in one of the left leads and a second particle at one of the
 right leads.  The ensuing 
the $4\times 4$ density matrix describes a system of two qubits with 
elements
\begin{eqnarray}\label{rho}
[\rho^{(2)}(\varepsilon)]_{ab,a'b'}&=&
\frac{1}{{\cal N}_0}\prod_{\alpha}\sum_{k_{\alpha}} \delta(\varepsilon-\epsilon_{k_{\alpha}})
\langle A_{ab}^\dagger A_{a'b'}  \rangle,\nonumber
\end{eqnarray}
where ${\cal N}_0$ is a normalization factor, while the mean value is taken in
the nonequilibrium state
with a net current flowing from the left to the right. 
We remark that in our approach matrix elements of $\rho^{(2)}$ are 
obtained in terms of operators appearing in the Hamiltonian $H$. Expectation values of four time dependent creation and/or annihilation operators
are computed using the nonequilibrium Green function formalism and Wick theorem, \cite{jauho}
\begin{widetext}
\begin{eqnarray} 
\langle c_{k_{\alpha}}^{\dagger}(t)c_{k_{\beta}}^{\dagger}(t)c_{k_{\lambda}}(t)c_{k_{\delta}}(t) \rangle
=\int_{-\infty}^{\infty} d\varepsilon \int_{-\infty}^{\infty} d\varepsilon^{\prime} \left[G_{k_{\lambda},k_{\alpha}}^{<}(\varepsilon-\varepsilon^{\prime})G_{k_{\delta},k_{\beta}}^{<}(\varepsilon^{\prime}) - G_{k_{\delta},k_{\alpha}}^{<}(\varepsilon-\varepsilon^{\prime})G_{k_{\lambda},k_{\beta}}^{<}(\varepsilon^{\prime})\right], \label{matrixelement}
\end{eqnarray}
\end{widetext}
where $G_{k_\alpha,k_\beta}^{<} (\varepsilon)$ is the Fourier transform with respect to $t-t^{\prime}$ of the lesser Green function 
\begin{equation}
G_{k_\alpha,k_\beta}^{<} (t-t^{\prime}) = i \langle c_{k_{\beta}}^{\dagger}(t^{\prime}) c_{k_{\alpha}}(t) \rangle .
\end{equation}
We  assume that $eV \sim 0$ and we are interested in analyzing $\varepsilon \sim \mu=(\mu_L + \mu_R)/2$. 
We present the corresponding expression of $\rho^{(2)}$ in Appendix A.

From the above density matrix we compute the concurrence, which is a 
good measure of entanglement. \cite{woot} In this case it is given by 
\begin{equation}
\mathcal{C}\left[\rho^{(2)} \right]=max\left\{0,\sqrt{\lambda_{1}}-
\sqrt{\lambda_{2}}-\sqrt{\lambda_{3}}-\sqrt{\lambda_{4}}\right\},
\end{equation}
where $\lambda_{i}$ are the eigenvalues of 
$R:=\rho^{(2)}\sigma_{y}\bigotimes\sigma_{y}\rho^{(2)\ast}\sigma_{y}\bigotimes\sigma_{y}$ in decreasing order.
It is interesting to notice that the concurrence so  calculated is equivalent to the one obtained by the spin--dependent scattering formalism 
for a single mode conductor ,\cite{been1,been2} which is
\begin{equation} 
\mathcal{C}=2\frac{\sqrt{\tau_{1}(1-\tau_{1})\tau_{2}(1-
\tau_{2})}}{\tau_{1}+\tau_{2}-\tau_{1}\tau_{2}}.
\end{equation}
Here, $\tau_{1},\tau_{2}$ are transmission eigenvalues of the scattering matrices 
$s_{\alpha,\beta}(\mu)=\delta_{\alpha,\beta}-
i\sqrt{\Gamma_{\beta} \Gamma_{\alpha}}G_{\alpha,\beta}^{R} (\mu)$,
where $G_{\alpha,\beta}^{R} (\mu)$ is the Fourier transform with respect to $t-t^{\prime}$ of the retarded Green function 
\begin{equation}\label{green}
G_{\alpha,\beta}^{R} (t-t^{\prime}) =- i \Theta(t-t^{\prime}) \sum_{\lambda,\lambda^{\prime}} \langle \{\Psi_{\lambda}( x_{\alpha},t),
\Psi^{\dagger}_{\lambda^{\prime}}( x_{\beta},t^{\prime}) \} \rangle 
\end{equation}
 evaluated at $\mu$, being $x_{\alpha}$ the position 
of the ring at which the wire $\alpha$ is attached. while $\Gamma_{\alpha} =2 \pi \sum_{k_{\alpha}} 
w_{k_{\alpha}}^2 \delta(\mu-\varepsilon_{k_{\alpha}}) $. \cite{fish-lee}
\
\subsection{Noise}

As pointed out in Refs. \onlinecite{been2, frust2} the concurrence 
can be expressed in terms of correlators that quantify the degree of 
violation of Bell inequalities. In transport setups the latter can 
in turn be directly related to current-current correlation functions, 
which are amenable to be experimentally detected. We, thus, 
turn to analyze the connection between these correlation functions 
and the above discussed entanglement. We first outline the procedure 
to compute the current-current noise within our treatment.
The current passing through the contact to the terminal $\alpha$, can be expressed by 
the following operator $ J_{\alpha} (t) = (e/ \hbar)  \sum_{k_{\alpha}}w_{k_{\alpha}} 
[i c_{k_{\alpha}}^{\dagger} (t) \Psi(x_{\alpha},t) + h.c] $.
The zero frequency shot-noise is a measure of 
the current-current correlations in different terminals. It reads
\begin{equation} \label{shotnoise}
{\cal S}_{\alpha,\beta}(0)=\frac{1}{2}\int d\tau \langle\{\delta J_{\alpha}(\tau),\delta J_{\beta}(0)\}\rangle,
\end{equation}
where $\delta J_{\alpha}=J_{\alpha}-\langle J_{\alpha}\rangle$. We calculate
(ec.(\ref{shotnoise})) by evaluating a bubble diagram in terms of nonequilibrium Green functions.
The corresponding expressions are presented in Appendix \ref{noise}.
\
\section{RESULTS}

\subsection{Qualitative Analysis}

\begin{figure}
 \centering
 \includegraphics[width=5cm]{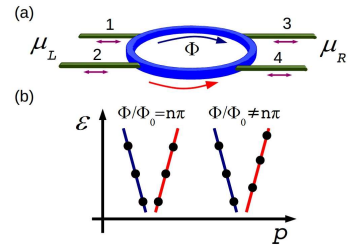}
 \caption{(a) Scheme of an Aharonov-Bohm ring with two chiralities, attached to four leads. The chemical potentials are such that there is a bias voltage between 
the left and right leads. Electrons incoming from the left can tunnel to the ring and escape through leads.  (b) Linear dispersion relation of an isolated AB-ring, where 
electrons move with velocity $v$ either clockwise ($+$ chirality) or anticlockwise ($-$ chirality). The magnetic flux determines whenever the Fermi level is two-degenerate or not.}
\label{fig1}
\end{figure}

To understand the origin of CME it is useful to begin
analyzing the electronic states of the isolated AB--ring. 
The Hamiltonian can be diagonalized in momentum space: Defining 
$\Psi_{\lambda}(x) = 1/\sqrt{{\cal N} } \sum_p e^{-i p x} c_{p,\lambda}$ 
with ${\cal N}$ a normalization factor, $p = 2 \pi n/L$ and 
$n \in \mathbf{Z}$ we obtain 
$H_{ring}=  \sum_{\lambda} \sum_p \varepsilon_{p,\lambda}(\Phi)  
c^{\dagger}_{p,\lambda} c_{p,\lambda} $, with 
$\varepsilon_{p,\lambda} = \lambda v (p - \phi) $ 
(a cutoff in the single-
particle energy spectrum is assumed). 
The effect of the magnetic flux on the energies 
$\varepsilon_{p,\lambda}(\Phi)$ is illustrated in Fig. \ref{fig1}. 
Depending on the magnetic flux, there can be zero, one 
or two single-particle states $|p,\lambda\rangle$ with 
a given energy. States with different chirality $\lambda = \pm $ 
are degenerate only when the flux is an integer multiple 
of $\pi \Phi_0$. 

Let us consider first the case where two degenerate states with 
opposite chiralities exist in the ring. These two states behaves as an intermediate qubit that couples to the qubits
defined by the leads. We will argue that in this case CME naturally emerges between electrons at the right and left leads. 
The $N$-particle states with a Fermi energy $\epsilon_F$ can be 
obtained from $|\overline{0} \rangle$, that represents the 
Fermi sea with $N-2$ particles filling the states with 
$\varepsilon_{p,\lambda}(\Phi) < \epsilon_F$,
where $v (p_F-\phi)=\epsilon_F$. Thus, we have 
$|\Psi_{ring}\rangle = c^{\dagger}_{p_F,+} c^{\dagger}_{p_F,-}|\overline{0}\rangle$.  
When the ring is in contact with the four leads, particles 
can tunnel between the ring and the reservoirs. 
For weak coupling 
we can assume that each of the  chiral levels with $p_F$ 
hybridize with the levels of the leads having the same 
energy $\epsilon_F$. This is described by the
following effective Hamiltonian
\begin{equation} \label{heff}
H_{eff}= w \sum_{\alpha=1}^4\sum_{\lambda=\pm} e^{i \lambda p_F x_{\alpha}}
[c^{\dagger}_{k_{\alpha}} c_{p_F,\lambda} + H. c.],
\end{equation}
where $w$ is the effective tunneling 
parameter, $c^{\dagger}_{k_{\alpha}}$ creates an electron 
in the single-particle state of the $\alpha$-lead
with energy $\varepsilon_{k_{\alpha}}=\epsilon_F$ (we take $\epsilon_F=0$ without
loss of generality). This Hamiltonian has four eigenstates of the 
form 
\begin{equation}
|\psi_n \rangle =[\sum_{\alpha=1}^4 \gamma_{n,\alpha} c^{\dagger}_{k_{\alpha}} + \sum_{\lambda=\pm} \gamma_{n,\lambda} c^{\dagger}_{p_F,\lambda} ]|0 \rangle, 
\;\;\;\; n=1, \ldots, 4,
\end{equation}
 where the coefficients $\gamma_{n}$ are the weights of the eigenstates in the chosen base. It also has two additional degenerate 
states of the form 
$|\psi_n \rangle =\sum_{\alpha=1}^4 \gamma_{n,\alpha} c^{\dagger}_{k_{\alpha}} |0 \rangle, \;n=5,6$. 
The latter correspond to states that do not hybridize with the ring.  
When two particles are present, two different such states must be 
occupied. It is simple to show that any state of this type 
has a sizable projection on states of the form $\sum_{\alpha \neq \beta}
\Lambda_{\alpha,\beta} c^{\dagger}_{k_{\alpha}} c^{\dagger}_{k_{\beta}}|0\rangle $ for 
some non-vanishing coefficients $\Lambda_{\alpha,\beta}$.
This two-particle state is typically entangled in the orbital 
indices $\alpha,\beta$ of opposite leads. 
Notice that it is also possible to use two AB--rings with degenerate levels as a two-qubit system coupled by two conducting leads (intermediate qubit). 

On the other hand if we consider the case of a one chirality ring ($\epsilon_{p}=v(p-\phi)$ for example) 
the situation drastically changes and no significant entanglement between left and right leads is attained. 
This is can be seen because the effective Hamiltonian has a different level structure. In fact,  
$H_{eff}$ can be naturally written  in terms of operators that 
are linear combinations of the lead operators $c^{\dagger}_{k_{\alpha}} $ as
$H_{eff}= w (f^{\dagger} c_{p_F,\lambda} + H. c.)$, with 
$f^{\dagger} = (1/2) \sum_{\alpha=1}^4 e^{i p_F x_{\alpha}} c^{\dagger}_{k_{\alpha}} $, while there are three additional orthogonal linear combinations
of these operators which do not hybridize with the ring.
The two eigenstates of $H_{eff}$ are linear combinations of a single-particle state of the leads and a single-particle state of the ring. 
Thus,
 a two-particle state of this Hamiltonian has never the two particles in the leads and therefore no orbital 
entanglement is possible.

The above argument suggests that by varying the magnetic flux, or the chemical potential of the leads (or equivalently, a gate voltage applied at the ring) 
we can induce the system to switch from a situation with no orbital entanglement between the leads 
onto another situation with inter-lead entanglement. This is done by varying $\Phi$ and/or $\mu$  
in order to have degenerate chiral states of the ring at the Fermi energy.
We now present a rigorous calculation of the entanglement for
the states that are relevant for a transport experiment in the 
coherent regime and explain how the CME depends on flux and chemical potential. We also discuss the way in which it may be detected in transport  experiments.

\subsection{Numerical Results}

\subsubsection{Concurrence}

\begin{figure}
 \includegraphics[width=9cm]{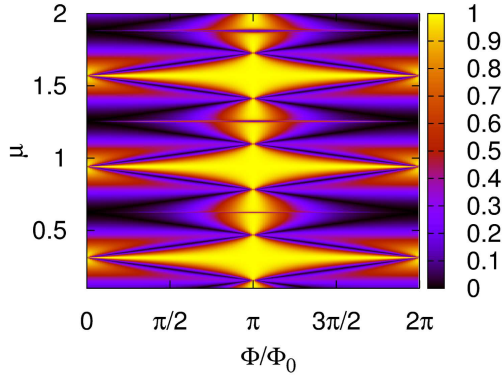}
 \caption{ (Color online) Concurrence  as a function of $\Phi / \Phi_0$ and mean chemical 
potential $\mu$ for a ring with length $L=20$ with wires connected at 
$x_1=1,\;x_2=6, x_3= 11, x_4= 16$.   }
\label{fig2}
\end{figure}

\begin{figure}
 \includegraphics[width=9cm]{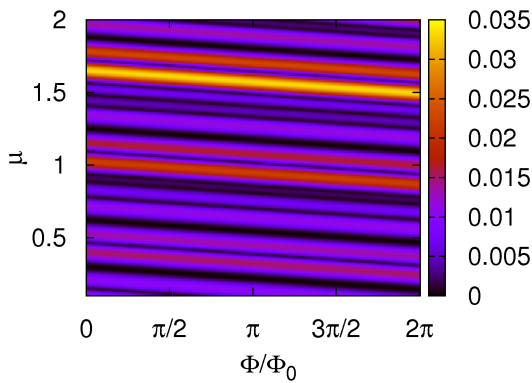}
 \caption{ (Color online) Concurrence  as a function of $\Phi / \Phi_0$ and mean chemical 
potential $\mu$ for a ring with only one chirality. The ring parameters are the same as in Fig.\ref{fig2}.}
\label{fig2b}
\end{figure}

The behavior of $ \mathcal{C}$ as a function of the mean chemical potential $\mu$ and the magnetic flux $\Phi$ is shown in Fig. \ref{fig2}. 
The concurrence is maximal at $\Phi/ \Phi_0=n \pi$ and for values $\mu$ close to the  energy of  two degenerate chiral states of the ring.  
The same type of behavior is observed for other configurations of the wires, corresponding
to contacts at different positions $x_{\alpha}$.  For this Fig. we considered  wires with a 
 bandwidth $W_{\alpha}$ and
$\Gamma_{\alpha}=  \Omega_{\alpha} \sqrt{ W_{\alpha}^2 - \mu^2}$, where $\Omega_{\alpha}$ is a constant, but the same behavior  is obtained for leads 
with a constant density of states. For some chemical potentials ${\cal C}$ exhibits maxima at $\Phi/\Phi_0= 0, \mod(2 \pi)$, which corresponds
to the energy of two degenerate states of the ring, but achieves again the maximum value within a wide range of fluxes centered at  $\Phi/\Phi_0= \pi, \mod(2 \pi)$.
This feature is analyzed below in more detail. 

Instead, if we evaluate ${\cal C}$ for the Hamiltonian (\ref{ring}) restricted to a single chirality,
we find negligibly orbital entanglement in the leads within the whole range of $\Phi$ and $\mu$. These results are presented in Fig.\ref{fig2b}. In this 
case the ring behaves as a single-level system, and prevent the formation of orbital entangled states at the leads.   

\begin{figure}
\centering
\includegraphics[width=8cm]{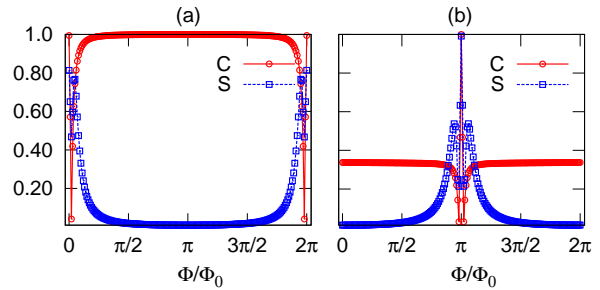}
\caption{(Color online)
Shot noise (squares) between left and right terminals, as a function of $\Phi/\Phi_0$  for two values of
the chemical potentials at which there are degenerate states.  (a) $\mu=0.3175$, (b) $\mu=0.7866$. The corresponding
plots for the concurrence are also shown for comparison (circles).}
\label{fig3}
\end{figure}

\subsubsection{Noise}

We now turn to present results on the behavior of the current-current correlation functions. Our aim is to analyze if the
signatures of entanglement found in the behavior of the concurrence can be also identified in the behavior of the noise.
Results for the total left-right noise 
correlations $S= \sum_{\alpha=1,2,\beta=3,4}{\cal S}_{\alpha,\beta}(0)$ 
(equal to the self-correlation $ - \sum_{\alpha,\beta=1,2}{\cal S}_{\alpha,\beta}(0)$ ) are shown in Fig. \ref{fig3}. 
The left (right) panel, corresponds to a chemical potential $\mu$ for which there are two degenerate chiral states in the ring for  
$\Phi/\Phi_0=0, (\pi), \mbox{mod}(2\pi)$. The behavior of the concurrence is also plotted for comparison. In both cases, 
 $S$, along with ${\cal C}$ exhibit maxima at the fluxes  for which the two degenerate 
chiral states are resonant at the given $\mu$.
 This points to the idea that noise is indeed a reliable witness of orbital
entanglement, as discussed in the context of other electronic setups. \cite{been1,been2,but1,noise-ent} In the case of the left panel,
both quantities are maximum at $\Phi/\Phi_0=0, \mbox{mod}(2\pi)$. Within a range of fluxes which scans
the width of the resonant degenerate levels of the ring they first decrease and then increase, displaying a dip. As 
the flux increases further, the behavior of these two quantities, however, depart one another. While $S$ tends to vanish around $\Phi/\Phi_0 = \pi$, the concurrence  displays a wide plateau
with height ${\cal C} \sim 1$. Qualitatively, the same type of
behavior is observed in the right panel.   In this case, both quantities exhibit a sharp maxima at resonance (see the peaks around $ \Phi/\Phi_0=\pi$). The  concurrence displays a plateau and another (lower) maximum around $\Phi/\Phi_0=0, \mbox{mod}(2 \pi)$ while 
$S$ is vanishing small. On general grounds this is rather surprising, since one could easily imagine situations with a sizable noise without entanglement, but here we have
the converse situation.

\begin{figure}
\centering
\includegraphics[width=8cm]{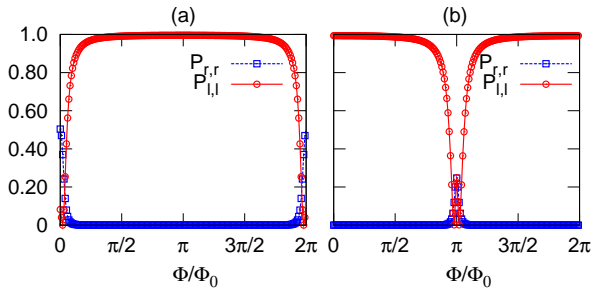}
\caption{(Color online) Probability of finding two particles at the left and right leads  as a function of magnetic flux $\Phi/\Phi_0$.
Panels (a) and (b) correspond to the same parameters of Fig. \ref{fig3} (a) and (b), respectively .}
\label{fig4}
\end{figure}

In order to further understand the relation between the behavior of ${\cal C}$ and $S$ as well as the connection to the CME,
we analyze the probabilities 
\begin{eqnarray}
P_{LL} &= &\prod_{\alpha}\sum_{k_{\alpha}} \delta(\varepsilon-\epsilon_{k_{\alpha}})\langle c^{\dagger}_{k_1}c_{k_1} c^{\dagger}_{k_2}c_{k_2} \rangle , \nonumber \\
P_{RR} &= &\prod_{\alpha}\sum_{k_{\alpha}} \delta(\varepsilon-\epsilon_{k_{\alpha}})\langle c^{\dagger}_{k_3}c_{k_3} c^{\dagger}_{k_4}c_{k_4} \rangle 
\end{eqnarray}
of finding two particles at the left (right)  leads, respectively. 
 In Fig. \ref{fig4} we show the  behavior of these two quantities for the same parameters of Fig. \ref{fig3}. 
It is clear that $P_{LL}$ and $P_{RR}$ are both sizable when tunneling through the ring is allowed. For these chemical potentials, this corresponds
to the narrow window  of fluxes around $\Phi/\Phi_0=0 (\pi)$, for the case of the left (right) panels, respectively,
within which the degenerate levels of the ring remain resonant. Beyond these values, the transmission from left to right is blocked and
the two particles have a high probability of remaining within the left wires. A further analysis of Figs. \ref {fig2} and \ref{fig3} in the light of 
the results shown in Fig. \ref{fig4}, then reveals that the high values of concurrence in the plateaus away from $\Phi/\Phi_0=0
(\pi), \mbox{mod}(2 \pi)$ in the case of the left (right) panels of Fig. \ref{fig3}, correspond to states 
 which have a very low probability of taking place.  Instead, the resonant situation
with two degenerate chiral states of the ring at the Fermi energy, leads to a high orbital entanglement which is clearly
witnessed by a high noise amplitude $S$.
\
\section{CONCLUSION}

To summarize, we introduced a new mechanism for orbital entanglement. This type of entanglement is originated in the
spectral nature of the AB ring and  is 
highly sensitive to the magnetic field.  Orbitally entangled electronic pairs can be produced by suitably tuning
the magnetic field, the chemical potential or a voltage gate at the ring, in order to have a degenerate pair of electronic states with different chiralities at the Fermi energy 
(intermediate qubit). In fact, if the right and left leads are mediated by a single--level system (single-chirality ring) the orbital entanglement
 is negligible.
This type of entanglement  can be detected in transport experiments, the shot noise being a good witness. 
The setup of Fig. \ref{fig1} could be experimentally realized in an architecture based on the quantum Hall regime of a 2D electron gas,
 by substituting each wire by a pair of incoming and outgoing edge states and the ring by a pair of edge states with different chiralities, separated by a narrow circular wall.
In such a setup it would  be possible to combine the main system of Fig. \ref{fig1} with beam splitters 
connected at the wires,
in order 
to test Bell inequalities and even perform the full quantum
tomography following the protocol of Ref. \onlinecite{but2}.
Two interesting possible generalizations are the combination between static and flying qubits
by combining with quantum dots, \cite{schom} as well as the introduction of dynamical single-particle emitters at the sources.\cite{spe} 
\
\section*{ACKNOWLEDGMENTS}

We thank D. Frustaglia for useful conversations.
This work is supported by CONICET, MINCyT and UBACYT, Argentina.
LA thanks support from the J. S. Guggenheim Foundation.
\
\appendix
\section*{APPENDIX}
\subsection*{A. Reduced Density Matrix} \label{aprho}
We evaluate the post-selected state of two electrons at opposite leads, $\rho^{(2)}$, in terms of Green functions (the Fourier transform of ec.(\ref{green})). 
The explicit expression of a general matrix element, up to a normalization factor, is:
\begin{widetext}
\begin{eqnarray}\label{rhoa}
&& \prod_{\nu=1}^{4}\sum_{k_{\nu}} \delta(\varepsilon-\epsilon_{k_{\nu}})\langle c_{k_{\alpha}}^{\dagger}(t)c_{k_{\beta}}^{\dagger}(t)c_{k_{\lambda}}(t)c_{k_{\delta}}(t) \rangle =\Gamma^{2}\left[f_{\alpha}(\varepsilon)f_{\beta}(\varepsilon) G_{_{\lambda},_{\alpha}}^{R}(\varepsilon)G_{_{\delta},_{\beta}}^{R}(\varepsilon)-f_{\alpha}(\varepsilon)f_{\delta}(\varepsilon) G_{_{\lambda},_{\alpha}}^{R}(\varepsilon)G_{_{\delta},_{\beta}}^{A}(\varepsilon) \right.\nonumber\\&&\left.
-f_{\lambda}(\varepsilon)f_{\beta}(\varepsilon) G_{_{\delta},_{\beta}}^{R}(\varepsilon)G_{_{\lambda},_{\alpha}}^{A}(\varepsilon)\right]-i\sum_{\gamma=1}^{4} \Gamma^{3}\left[ f_{\beta}(\varepsilon)f_{\gamma}(\varepsilon) G_{_{\lambda},_{\gamma}}^{R}(\varepsilon)G_{_{\delta},_{\beta}}^{R}(\varepsilon)G_{_{\gamma},_{\alpha}}^{A}(\varepsilon) \right.\nonumber\\&&\left.
f_{\alpha}(\varepsilon)f_{\gamma}(\varepsilon) G_{_{\lambda},_{\alpha}}^{R}(\varepsilon)G_{_{\delta},_{\gamma}}^{R}(\varepsilon)G_{_{\gamma},_{\beta}}^{A}(\varepsilon)\right] 
+\sum_{\gamma,\eta=1}^{4}\Gamma^{4}f_{\eta}(\varepsilon)f_{\gamma}(\varepsilon) G_{_{\lambda},_{\gamma}}^{R}(\varepsilon)G_{_{\delta},_{\eta}}^{R}(\varepsilon)G_{_{\eta},_{\beta}}^{A}(\varepsilon)G_{_{\gamma},_{\alpha}}^{A}(\varepsilon)\nonumber\\
&&+ \Gamma \left[-i f_{\alpha}(\varepsilon)f_{\beta}(\varepsilon) G_{_{\delta},_{\beta}}^{R}(\varepsilon)+if_{\alpha}(\varepsilon)f_{\delta}(\varepsilon) G_{_{\delta},_{\beta}}^{A}(\varepsilon)-\Gamma\sum_{\gamma=1}^{4} f_{\gamma}(\varepsilon)f_{\alpha}(\varepsilon)G_{_{\delta},_{\gamma}}^{R}(\varepsilon)G_{_{\gamma},_{\beta}}^{A}(\varepsilon) \right]\delta_{k_{\alpha},k_{\lambda}} \nonumber\\&&
+\Gamma \left[-i f_{\delta}(\varepsilon)f_{\alpha}(\varepsilon) G_{_{\lambda},_{\alpha}}^{R}(\varepsilon)+if_{\lambda}(\varepsilon)f_{\delta}(\varepsilon) G_{_{\lambda},_{\alpha}}^{A}(\varepsilon)-\Gamma\sum_{\gamma=1}^{4} f_{\gamma}(\varepsilon)f_{\delta}(\varepsilon)G_{_{\lambda},_{\gamma}}^{R}(\varepsilon)G_{_{\gamma},_{\alpha}}^{A}(\varepsilon) \right]\delta_{k_{\delta},k_{\beta}}\nonumber\\&&
-\sum_{k_{\alpha}}\sum_{k_{\delta}} \delta(\varepsilon-\epsilon_{k_{\alpha}})\delta(\varepsilon-\epsilon_{k_{\delta}})f_{\delta}(\varepsilon)f_{\alpha}(\varepsilon)\delta_{k_{\alpha},k_{\lambda}}\delta_{k_{\delta},k_{\beta}},
\end{eqnarray}
\end{widetext}
where we considered all spectral densities equal $\Gamma_{\alpha}=\Gamma$ $\forall \alpha$. $f_{\alpha}(\varepsilon)$ is de Fermi distribution 
function of the lead $\alpha$.  Note that $G_{\alpha,\beta}^{R}(\varepsilon)=\sum_{\lambda,\lambda^{\prime}}G_{\lambda,\lambda^{\prime}}^{R}(x_{\alpha},x_{\beta},\varepsilon)$.The retarded Green functions are evaluated by solving  the Dyson equation,
\begin{widetext}  
\begin{eqnarray}
G^{R}_{\lambda,\lambda^{\prime}}(x,x^{\prime},\varepsilon)= g^{R}_{\lambda^{\prime}}(x,x^{\prime},\varepsilon)\delta(x-x^{\prime})\delta_{\lambda,\lambda^{\prime}}+\sum_{\gamma=1}^{4}\sum_{\lambda^{\prime\prime}}G^{R}_{\lambda,\lambda^{\prime\prime}}(x,x_{\gamma},\varepsilon)\Sigma_{\gamma}^{R}(\varepsilon)g^{R}_{\lambda^{\prime}}(x_{\gamma},x^{\prime},\varepsilon),
\end{eqnarray}
\end{widetext}  
being 
\begin{equation}
g^{R}_{\lambda^{\prime}}(x,x^{\prime},\varepsilon)=\frac{1}{\mathcal{M}}\sum_{k=-k_{0}}^{k_{0}}\frac{e^{-ik(x-x^{\prime})}}{\varepsilon-\varepsilon_{k,\lambda^{\prime}}(\Phi)+i\eta},
\end{equation} 
where $k_{0}$ is the energy cut off and $\mathcal{M}=2k_{0}+1$ a normalization factor,  
the uncoupled retarded Green function of the ring and $\Sigma_{\gamma}^{R}(\varepsilon)=\sum_{k_{\alpha}}|w_{k_{\alpha}}|^{2}g_{k_{\alpha}}^{R}(\varepsilon)$ the 
self energy of reservoir $\alpha$.

\subsection*{B. Shot Noise Calculation}\label{noise}
Here we show the shot noise expression of ec.(\ref{shotnoise}) in terms of retarded Green functions. \cite{jauho}
\begin{widetext}  
\begin{eqnarray}&S_{\alpha,\beta}(0)&=\frac{1}{2}\int d\tau \left[ \langle\{J_{\alpha}(\tau),J_{\beta}(0)\}\rangle - 2\langle J_{\alpha}(\tau)\rangle \langle J_{\beta}(0)\rangle \right] \\ \nonumber
&& = \frac{e^{2}}{2\hbar} \int \frac{d\varepsilon}{2\pi} \lgroup \sum_{\delta,\gamma=1}^{4}\left[\Gamma_{\alpha}\Gamma_{\beta}\Gamma_{\delta}\Gamma_{\gamma} [(1-f_{\gamma}(\varepsilon))f_{\delta}(\varepsilon)+(1-f_{\delta}(\varepsilon))f_{\gamma}(\varepsilon)]G_{\beta,\gamma}^{R}(\varepsilon)G_{\alpha,\gamma}^{R*}(\varepsilon)G_{\alpha,\delta}^{R}(\varepsilon)G_{\beta,\delta}^{R*}(\varepsilon)\right] \\ \nonumber   
&& -2\Gamma_{\alpha}\Gamma_{\beta}[(1-f_{\beta}(\varepsilon))f_{\alpha}(\varepsilon)+(1-f_{\alpha}(\varepsilon))f_{\beta}(\varepsilon)]\mathcal{R}e \{ G_{\alpha,\beta}^{R}(\varepsilon)G_{\beta,\alpha}^{R}(\varepsilon) \} \\ \nonumber
&& -2\Gamma_{\alpha}\Gamma_{\beta}\mathcal{R}e \{ iG_{\beta,\alpha}^{R}(\varepsilon)\sum_{\delta=1}^{4}G_{\alpha,\delta}^{R}(\varepsilon)G_{\beta,\delta}^{R*}(\varepsilon)\Gamma_{\delta}\left[(1-f_{\alpha}(\varepsilon))f_{\delta}(\varepsilon)+(1-f_{\delta}(\varepsilon))f_{\alpha}(\varepsilon)\right]\} \\ \nonumber
&& -2\Gamma_{\alpha}\Gamma_{\beta}\mathcal{R}e \{ iG_{\alpha,\beta}^{R}(\varepsilon)\sum_{\delta=1}^{4}G_{\beta,\delta}^{R}(\varepsilon)G_{\alpha,\delta}^{R*}(\varepsilon)\Gamma_{\delta}\left[(1-f_{\beta}(\varepsilon))f_{\delta}(\varepsilon)+(1-f_{\delta}(\varepsilon))f_{\beta}(\varepsilon)\right]\}\\ \nonumber
&& + 2\delta_{\alpha,\beta} \Gamma_{\alpha}\sum_{\delta=1}^{4} \Gamma_{\delta}|G_{\alpha,\delta}^{R}(\varepsilon)|^{2}\left[(1-f_{\delta}(\varepsilon))f_{\alpha}(\varepsilon)+(1-f_{\alpha}(\varepsilon))f_{\delta}(\varepsilon)\right]\rgroup.
\end{eqnarray}
\end{widetext}



\begin{thebibliography}{11}

\bibitem{qe} R. Horodecki {\it et al}, Rev. Mod. Phys. {\bf 81}, 865-942 (2009).

\bibitem{inter} R. Hanbury Brown and R. Q. Twiss, Nature {\bf 178} , 1046 (1956);
C. K. Hong, Z. Y. Ou, and L. Mandel, Phys. Rev. Lett. {\bf 59}, 2044 (1987).

\bibitem{entqdot} D. Loss and D. P. DiVincenzo, Phys. Rev. A {\bf 57}, 120 (1998);
R. Hanson,  L .P. Kouwenhoven, J. R. Petta, S. Tarucha, and L. M. K. Vandersypen, Rev. Mod. Phys. {\bf 79}, 1217 (2007).

\bibitem{entsup} J. Wei, and V. Chandrasekhar, Nature Phys. 6, 494 (2010).

\bibitem{been1} C. W. J. Beenakker, C. Emary, M. Kindermann, and J. L. van Velsen, 
Phys. Rev. Lett. {\bf 91},
147901 (2003).

\bibitem{been2}C. W. J. Beenakker, in {\em ''Quantum Computers, Algorithms and
Chaos''}, Proceedings of the International School of Physics E. Fermi,  Varenna, 
2005,
(IOS Press, Amsterdam, 2006); C. W. J. Beenakker, M. Kindermann, C. M. Marcus, and
A. Yacoby in {\em ''Fundamental Problems of Mesoscopic Physics''}, edited by
I. V. Lerner, B. L. Altshuler and Y. Gefen, NATO Science Series II, vol 154 
(Kluwer, Dordrecht, 2004).

\bibitem{elin} M. Henny,  
S. Oberholzer, C. Strunk, T. Heinzel, K. Ensslin, M. Holland, and C. Sch\"onenberger; Science 
{\bf 284}, 296 (1999); W.D. Oliver, J. Kim, R.C. Liu, and Y. Yamamoto, ibid. {\bf 284}, 299 (1999); 
 H. Kiesel, A. Renz,
and F. Hasselbach, Nature  {\bf 418}, 392 (2002); Y. Ji, Y. Chung, D. Sprinzak, M. Heiblum, D. Mahalu,
and H. Shtrikman, Nature {\bf 422}, 415 (2003);
I. Neder, N. Ofek, Y. Chung, M. Heiblum, D. Mahalu, and
V. Umansky, Nature  {\bf 448}, 333 (2007); V. Giovannetti, D. Frustaglia, F. Taddei, and R. Fazio,
Phys. Rev. B {\bf 74}, 115315 (2006). 

\bibitem{hbt}I. Neder, N. Ofek, Y. Chung, M. Heiblum, D. Mahalu, V. Umansky, Nature  {\bf 448}, 333 (2007).

\bibitem{but1}P. Samuelsson, E.V. Sukhorukov, M. B\"uttiker, Phys. Rev. Lett. {\bf 
92}, 026805 (2004).

\bibitem{but2}P. Samuelsson, M. B\"uttiker, Phys. Rev. B {\bf 73}, 041305 (2006).

\bibitem{noise-ent} I. Klich and L. Levitov, Phys. Rev. Lett. {\bf 102}, 100502 (2009);
B. Hsu, E. Grosfeld and  E. Fradkin
 Phys. Rev. B {\bf 80}, 235412 (2009).

\bibitem{frust1}V. A. Gopar, D. Frustaglia, Phys. Rev. B {\bf 77}, 153403 (2008).

\bibitem{frust2}D Frustaglia and A. Cabello, Phys. Rev. B {\bf 80}, 201312(R) (2009).

\bibitem{woot} W. K. Wooters, Phys. Rev. Lett. {\bf 80}, 2245 (1998).

\bibitem{jauho} H. J. W. Haug and A-P. Jauho, ''Quantum kinetics in transport and optics of semiconductors'', Springer (2008).

\bibitem{fish-lee} D. S. Fisher and P. A. Lee,  Phys. Rev. B {\bf 23}, 6851 (1981).

\bibitem{schom} H. Schomerus, and J. P. Robinson, New J. Phys {\bf 9}, 67 (2007).

\bibitem{spe}G. F\`eve, A. Mahe, J.-M. Berroir, T. Kontos, B. Placais, DC Glattli, A. Cavanna, B. Etienne, Y. Jin, Science {\bf 316}, 1169 (2007);  J. Splettstoesser, M. Moskalets, M. B\"uttiker
 Phys. Rev. Lett. {\bf 103}, 076804 (2009).




\end{thebibliography}
\end{document}